-----------------------------------------------------------
\documentstyle[11pt]{article}
\begin{document}
$~~~$

\vspace{15mm}
\noindent
\parbox[]{4.7in}{{\large \bf EXACT SOLUTION TO THE SCHR\"{O}DINGER}\\[1.5mm]
{\large \bf EQUATION FOR THE QUANTUM RIGID BODY}}

\vspace{6mm}
\noindent
\hspace{15mm}{\bf ZHONG-QI MA}

\vspace{3mm}
\noindent
\hspace{15mm}\parbox[]{4in}{{\it Department of College 
Computer Education, Hu'nan Normal University, Changsha 410081,
The People's Republic of China, and Institute of High Energy 
Physics, Beijing 100039, The People's Republic of China.}}

\vspace{6mm}
\noindent
The exact solution to the Schr\"{o}dinger equation for the 
rigid body with the given angular momentum and parity is 
obtained. Since the quantum rigid body can be thought of as 
the simplest quantum three-body problem where the internal 
motion is frozen, this calculation method is a good starting 
point for solving the quantum three-body problems.

\vspace{4mm}

\vspace{6mm}
\noindent
{\bf 1. INTRODUCTION}

\vspace{2mm}
\noindent
The three-body problem is a fundamental problem in quantum 
mechanics, which has not been well solved. The Faddeev equations 
[1] provide a method for solving exactly the quantum three-body 
problems. However, only a few analytically solvable examples were 
found [2]. The accurate direct solution of the three-body 
Schr\"{o}dinger equation with the separated center-of-mass motion 
has been sought based on different numerical methods, such as the 
finite difference [3], finite element [4], complex coordinate 
rotation [5], hyperspherical coordinate [6-8], hyperspherical 
harmonic [9-11] methods, and a large number of works [12-16]. In those 
numerical methods, three rotational degrees of freedom are not 
separated completely from the internal ones. In this letter we 
present a method to separate completely the rotational degrees of 
freedom and apply it to the quantum rigid body as an example. 

The plan of this letter is organized as follows. In Sec. 2 we
shall introduce our notations and briefly demonstrate how to 
separate the rotational degrees of freedom from the internal 
ones in a quantum three-body problem. The exact solution to 
the Schr\"{o}dinger equation for the rigid body with the given 
angular momentum and parity is obtained in Sec. 3. A short
 conclusion is given in sec. 4.

\vspace{4mm}
\noindent
{\bf 2. QUANTUM THREE-BODY PROBLEM}

\vspace{2mm}
\noindent
Denote by ${\bf r_{j}}$ and by $M_{j}$, $j=1,2,3$, 
the position vectors and the masses of three particles in a 
three-body problem, respectively. The relative masses are 
$m_{j}=M_{j}/M$, where $M$ is the total mass, $M=\sum M_{j}$. 
The Laplace operator in the three-body Schr\"{o}dinger equation 
is proportional to $\sum_{j=1}^{3}~m_{j}^{-1}\bigtriangleup_{{\bf r}_{j}}$, 
where $\bigtriangleup_{{\bf r}_{j}}$ is the Laplace operator 
with respect to the position vector ${\bf r}_{j}$.
Introducing the Jacobi coordinate vectors ${\bf x}$ and ${\bf y}$ 
in the center-of-mass frame,
$${\bf x}=-\sqrt{\displaystyle {m_{1}\over m_{2}+m_{3}}}~{\bf r_{1}},~~~~~
{\bf y}=\sqrt{\displaystyle {m_{2}m_{3}\over m_{2}+m_{3}}}
~\left({\bf r_{2}-r_{3}}\right).  \eqno (1) $$

\noindent
we obtain the Laplace operator and the total angular momentum 
operator ${\bf L}$ by a direct replacement of variables:
$$\begin{array}{c}
\bigtriangleup = \displaystyle \sum_{j=1}^{3}~
m_{j}^{-1} \bigtriangleup_{{\bf r}_{j}}  
= \bigtriangleup_{{\bf x}}+ \bigtriangleup_{{\bf y}},\\
{\bf L}= \displaystyle \sum_{j=1}^{3}~
-i\hbar {\bf r}_{j}\times \bigtriangledown_{{\bf r}_{j}}
={\bf L}_{{\bf x}}+{\bf L}_{{\bf y}},\\
{\bf L}_{{\bf x}}=-i\hbar {\bf x}\times \bigtriangledown_{{\bf x}},~~~~~
{\bf L}_{{\bf y}}=-i\hbar {\bf y}\times \bigtriangledown_{{\bf y}}. 
\end{array} \eqno (2) $$

\noindent
The three-body Schr\"{o}dinger equation with the separated 
center-of-mass motion becomes 
$$- \left(\hbar^{2}/2M\right) \left\{
\bigtriangleup_{{\bf x}}+\bigtriangleup_{{\bf y}}\right\} \Psi 
+V \Psi =E \Psi, \eqno (3) $$

\noindent
where $V$ is a pair potential, depending only upon the distance 
of each pair of particles.

In the hyperspherical harmonic method [11], for example, two
Jacobi coordinate vectors are expressed in their spherical
coordinate forms, 
$${\bf x}\sim (\rho \cos \omega, \theta_{x}, \varphi_{x}),~~~~~
{\bf y}\sim (\rho \sin \omega, \theta_{y}, \varphi_{y}). \eqno (4) $$

\noindent
where $\rho$ is called the hyperradius and 
$\Omega(\omega, \theta_{x}, \varphi_{x},\theta_{y}, \varphi_{y})$
are the five hyperangular variables. The wave function is 
presented as a sum of products of a hyperradial function and 
the hyperspherical harmonic function,
$$\Psi_{\ell m}({\bf x,y})=\displaystyle \sum_{K,\ell_{x} \ell_{y}} 
\psi_{K,\ell_{x} \ell_{y}}(\rho) {\cal Y}^{\ell m}_{K,\ell_{x} \ell_{y}}
(\Omega). $$

\noindent
There is huge degeneracy of the hyperspherical basis, and the 
matrix elements of the potential have to be calculated between 
different hyperspherical harmonic states [10], because the 
interaction in the three-body problem is not hyperspherically
symmetric.

The quantum rigid body (top) can be thought of as the simplest quantum
three-body problem where the internal motion is frozen. To 
solve exactly the Schr\"{o}dinger equation for the rigid body is the 
first step for solving exactly the quantum three-body problems. 
Wigner first studied the exact solution for the quantum
rigid body (see P.214 in [17]) from the group theory. He 
characterized the position of the rigid body by the three Euler angles 
$\alpha$, $\beta$, $\gamma$ of the rotation which brings the rigid body 
from its normal position into the position in question, and 
obtained the exact solution for the quantum rigid body, which is
nothing but the Wigner $D$-function. For the quantum three-body
problems, as in the helium atom, he separated three rotational
degrees of freedom from three internal ones
(see Eq. (19.18) in [17]):
$$\Psi_{\ell m}({\bf r_{1},r_{2}})=\displaystyle \sum_{\nu}~
D^{\ell}_{m \nu}(\alpha,\beta,\gamma)^{*}
\psi_{\nu}(r_{1},r_{2},\omega),
\eqno (5) $$

\noindent
where ${\bf r}_{1}$ and ${\bf r}_{2}$ are the coordinate vectors
of two electrons, $\omega$ is their angle,
and the Wigner $D$-function form [17] has been replaced with
the usual $D$-function form [18]. Wigner did not write the 
three-body Schr\"{o}dinger equation explicitly. As a matter of 
fact, the three-body Schr\"{o}dinger equation (3) 
becomes very complicated if one replaces two coordinates vectors 
of electrons with the Euler angles as well as $r_{1}$, $r_{2}$, 
and $\omega$ for the internal motion. 
On the other hand, Wigner's idea, to separate the degrees 
of freedom completely from the internal ones, is helpful to simplify 
the calculation for the quantum three-body problem. Hsiang and Hsiang 
in their recent paper [19] also presented the similar idea.
In this letter we will develop the idea of Wigner and obtain 
the exact solution of the Schr\"{o}dinger equation for the 
rigid body without introducing the Euler angles directly. 
This calculation method is a good starting point for solving 
the quantum three-body problems [19,20].

The Schr\"{o}dinger equation (3) is spherically symmetric so 
that its solution can be factorized into a product of an
eigenfunction of the angular momentum ${\bf L}$ and a
"radial" function, which only depends upon three variables,
invariant in the rotation of the system:
$$\xi_{1}={\bf x\cdot x},~~~~~
\xi_{2}={\bf y\cdot y},~~~~~
\xi_{3}={\bf x\cdot y}. \eqno (6) $$

\noindent
For the quantum rigid body, the potential makes the internal
motion frozen so that those variables $\xi_{j}$ are constant.

For a particle moving in a central field, the eigenfunction of 
the angular momentum is the spherical harmonic function 
$Y^{\ell}_{m}(\theta, \varphi)$. How to generalize
the spherical harmonic function to the three-body problem
without introducing the Euler angles directly? As is 
well known, 
${\cal Y}^{\ell}_{m}({\bf x})= r^{\ell}Y^{\ell}_{m}(\theta, \varphi)$,
where $(r,\theta,\varphi)$ are the spherical coordinates for
the position vector ${\bf x}$, is a homogeneous polynomial of 
degree $\ell$ with respect to the components of ${\bf x}$,
which does not contain $r^{2}={\bf x \cdot x}$ explicitly. 
${\cal Y}^{\ell}_{m}({\bf x})$, called the harmonic polynomial 
in the literature, satisfies the Laplace equation 
as well as the eigen-equation for the angular momentum.

In the three-body problem there are two Jacobi coordinate
vectors ${\bf x}$ and ${\bf y}$ in the center-of-mass frame. 
We shall construct the eigenfunctions of the angular momentum
as the homogeneous polynomials of degree
$\ell$ with respect to the components of ${\bf x}$ and ${\bf y}$, 
which do not contain $\xi_{j}$ explicitly. According to the theory of 
angular momentum [18], they are
$$\begin{array}{c}
{\cal Y}^{\ell q}_{L m}({\bf x,y})=\displaystyle \sum_{\mu}~
{\cal Y}^{q}_{\mu}({\bf x}){\cal Y}^{\ell-q}_{m-\mu}({\bf y})
\langle q,\mu,\ell-q,m-\mu|q,\ell-q,L,m\rangle , \\
0\leq q \leq \ell,~~~{\rm when}~~ L=\ell, ~~{\rm and}~~ 
1\leq q \leq \ell-1, ~~~{\rm when}~~ L=\ell-1. 
\end{array} \eqno (7) $$

\noindent
where $\langle q,\mu,\ell-q,m-\mu|q,\ell-q,L,m\rangle$
are the Clebsch-Gordan coefficients. 
The remained combinations with the angular momentum $L<\ell-1$ 
contain the factors $\xi_{3}$ explicitly [20]. In other words, 
the eigenfunctions of the total angular momentum ${\bf L}^{2}$
with the eigenvalue $\ell(\ell+1)$, not containing the factors 
$\xi_{j}$ explicitly, are those homogeneous polynomials of 
degree $\ell$ or degree $(\ell+1)$. Let us introduce a parameter 
$\lambda=0$ or $1$ to identify them:
$$\begin{array}{rl}
{\cal Y}^{(\ell+\lambda)q}_{\ell m}({\bf x,y})
&=~\displaystyle \sum_{\mu}
{\cal Y}^{q}_{\mu}({\bf x}){\cal Y}^{\ell-q+\lambda}_{m-\mu}({\bf y})\\
&~~~\times~\langle q,\mu,\ell-q+\lambda,m-\mu|q,\ell-q+\lambda,\ell,m
\rangle , \\[2mm]
&\lambda=0~~{\rm and}~~1,~~~~~~~\lambda \leq q \leq \ell. 
\end{array} \eqno (8) $$

${\cal Y}^{(\ell+\lambda)q}_{\ell m}({\bf x,y})$ is the common eigenfunction
of ${\bf L}^{2}$, $L_{3}$, ${\bf L}_{{\bf x}}^{2}$, ${\bf L}_{{\bf y}}^{2}$, 
$\bigtriangleup_{{\bf x}}$, $\bigtriangleup_{{\bf y}}$, 
$\bigtriangleup_{{\bf xy}}$, and the parity with the eigenvalues 
$\ell(\ell+1)$, $m$, $q(q+1)$, $(\ell-q+\lambda)(\ell-q+\lambda+1)$, $0$, $0$, $0$, 
and $(-1)^{\ell+\lambda}$, respectively, 
where ${\bf L}^{2}$ and $L_{3}$ are the total angular 
momentum operators, ${\bf L}_{{\bf x}}^{2}$ and ${\bf L}_{{\bf y}}^{2}$ are 
the "partial" angular momentum operators [see Eq. (2)], 
$\bigtriangleup_{{\bf x}}$ and $\bigtriangleup_{{\bf y}}$ are 
the Laplace operators respectively with respect to the Jacobi coordinate
vectors ${\bf x}$ and ${\bf y}$, and $\bigtriangleup_{{\bf xy}}$ 
is defined as
$$\bigtriangleup_{{\bf xy}}=\displaystyle {\partial^{2} \over 
\partial x_{1}\partial y_{1} }+ \displaystyle {\partial^{2} \over 
\partial x_{2}\partial y_{2} } +\displaystyle {\partial^{2} \over 
\partial x_{3}\partial y_{3} }. \eqno (9) $$

Because of the conservation of the angular momentum and parity, 
the solution $\Psi_{\ell m \lambda}({\bf x,y})$ of the 
Schr\"{o}dinger equation (3) can be expanded in terms of 
${\cal Y}^{(\ell+\lambda)q}_{\ell m}({\bf x,y})$, where the
conserved quantum numbers $\ell$, $m$ and $\lambda$ are fixed. 
Since those equations are independent of $m$, we can calculate
them by setting $m=\ell$, where [18]  
$$\begin{array}{l}
{\cal Y}^{\ell q}_{\ell \ell}({\bf x,y})\\
~~~=(-1)^{\ell} \left\{ \displaystyle 
{ \left[(2q+1)!(2\ell-2q+1)!\right]^{1/2} \over q! (\ell-q)! 2^{\ell+2}\pi}
\right\}(x_{1}+i x_{2})^{q}(y_{1}+iy_{2})^{\ell-q}, \\
{\cal Y}^{(\ell+1)q}_{\ell \ell}({\bf x,y})\\
~~~=(-1)^{\ell} 
\left\{ \displaystyle 
{ (2q+1)!(2\ell-2q+3)! \over 2q(\ell-q+1)(\ell+1)} \right\}^{1/2}
\left\{(q-1)! (\ell-q)! 2^{\ell+2}\pi \right\}^{-1}\\[3mm]
~~~~~~\times~(x_{1}+i x_{2})^{q-1}(y_{1}+iy_{2})^{\ell-q} 
\left\{(x_{1}+ix_{2})y_{3}-
x_{3}(y_{1}+iy_{2})\right\}^{\lambda} .
\end{array}  \eqno (10) $$ 

\noindent
By substituting $\Psi_{\ell \ell \lambda}({\bf x,y})$ into Eq. (3), 
a system of the partial differential equations for the coefficients
can be obtained. The partial differential equations will be 
simplified if one changes the normalization factor of 
${\cal Y}^{(\ell+\lambda)q}_{\ell \ell}({\bf x,y})$,
namely ${\cal Y}^{(\ell+\lambda)q}_{\ell \ell}({\bf x,y})$ in Eq. (11) is
replaced by $Q^{\ell \lambda}_{q}({\bf x,y})$, which is proportional to
${\cal Y}^{(\ell+\lambda)q}_{\ell \ell}({\bf x,y})$:
$$\begin{array}{rl}
\Psi_{\ell\ell \lambda}({\bf x,y})&=~\displaystyle \sum_{q=\lambda}^{\ell}~
\psi_{q}^{\ell \lambda}(\xi_{1},\xi_{2},\xi_{3})
Q_{q}^{\ell \lambda}({\bf x,y}),\\[2mm]
Q^{\ell\lambda}_{q}({\bf x,y})&=~
\left\{(q-\lambda)! (\ell-q)! \right\}^{-1}
(x_{1}+i x_{2})^{q-\lambda}(y_{1}+iy_{2})^{\ell-q} \\[2mm]
&~~~\times~\left\{(x_{1}+ix_{2})y_{3}-
x_{3}(y_{1}+iy_{2})\right\}^{\lambda} \\[2mm]
&\lambda=0,1,~~~~~~\lambda \leq q \leq \ell. 
\end{array} \eqno (11) $$

\noindent
The partial differential equations for the functions 
$\psi_{q}^{\ell \lambda}(\xi_{1},\xi_{2},\xi_{3})$ are:
$$\begin{array}{l}
\displaystyle -{\hbar^{2} \over 2M} \left\{ \bigtriangleup
\psi^{\ell \lambda}_{q} +4q \displaystyle 
{\partial \psi^{\ell \lambda}_{q} \over \partial \xi_{1}}
+4(\ell-q+\lambda) \displaystyle 
{\partial \psi^{\ell \lambda}_{q} \over \partial \xi_{2}}
+2(q-\lambda) \displaystyle 
{\partial \psi^{\ell \lambda}_{q-1} \over \partial \xi_{3}} \right. \\
~~~~~~\left.+2(\ell-q) \displaystyle 
{\partial \psi^{\ell \lambda}_{q+1} \over \partial \xi_{3}} \right\} 
=(E-V) \psi^{\ell \lambda}_{q},\\
~~~~~~~~~~~~~\lambda\leq q \leq \ell,~~~~~\lambda =0,1. 
\end{array} \eqno (12) $$

\noindent
This system of the partial differential equations was first 
obtained by Hsiang and Hsiang [19]. It is a good starting point 
for solving the quantum three-body problems [19,20].

\vspace{4mm}
\noindent
{\bf 3. QUANTUM RIGID BODY}

\vspace{2mm}
\noindent
For the quantum rigid body, the potential preserves the 
geometrical form of the rigid body fixed. It can be replaced 
by the constraints:
$$\xi_{1}={\rm const}.~~~~~\xi_{2}={\rm const}.~~~~~\xi_{3}={\rm const}.
\eqno (13) $$

\noindent
Therefore, the solution of the Schr\"{o}dinger equation for the 
quantum rigid body can be expressed as
$$\Psi_{\ell\ell \lambda}({\bf x,y})=\displaystyle \sum_{q=\lambda}^{\ell}~
f_{q}^{\ell \lambda}Q_{q}^{\ell \lambda}({\bf x,y}). \eqno (14) $$

\noindent
where $f_{q}^{\ell \lambda}$ are constant. 
Recall that $Q^{\ell\lambda}_{q}({\bf x,y})$ is the solution of the 
Laplace equation. Due to the constraints (13) some differential
terms with respect to $\xi_{j}$ in the Laplace equation 
should be removed so that the Laplace equation is violated,
namely, the rigid body obtains an energy $E$. On the other hand, 
as a technique of calculation, we can calculate those differential 
terms first where $\xi_{j}$ are not constant, and then set the 
constraints (13). The contribution from those terms is nothing 
but the minus energy $-E$ of the rigid body. 

In the calculation, we first separate the six Jacobi coordinates
[see Eq. (4)] into three rotational coordinates and three internal 
coordinates. The lengths of ${\bf x}$ and ${\bf y}$ and their 
angle $\omega$ are
$$r_{x}=\sqrt{\xi_{1}},~~~~~r_{y}=\sqrt{\xi_{2}},~~~~~
\cos \omega=\xi_{3}/\sqrt{\xi_{1}\xi_{2}}. \eqno (15) $$

\noindent
Obviously, those three variables are also constant in the 
constraints (13). Assume that in the normal position of the 
rigid body the Jacobi coordinate vector ${\bf x}$ is along 
the $Z$ axis and ${\bf y}$ is located in the $XZ$ plane with 
a positive $X$ component. A rotation $R(\alpha,\beta,\gamma)$
brings the rigid body from its normal position into the position 
in question. The Euler angles $\alpha$, $\beta$, and $\gamma$ 
describe the rotation of the rigid body. The definition for
the Euler angles are different from that of Wigner 
(see Eq. (7) and Ref. [17]) because ${\bf x}$ and ${\bf y}$ here
are the Jacobi coordinate vectors. To shorten the notations, we define
$$\begin{array}{lll}
c_{\alpha}=\cos \alpha,~~~~~c_{\beta}=\cos \beta,~~~~~
c_{\gamma}=\cos \gamma,\\
c_{x}=\cos \theta_{x},~~~~~c_{y}=\cos \theta_{y},~~~~~
C=\cos \omega,\\
s_{\alpha}=\sin \alpha,~~~~~s_{\beta}=\sin \beta,~~~~~
s_{\gamma}=\sin \gamma,\\
s_{x}=\sin \theta_{x},~~~~~s_{y}=\sin \theta_{y},~~~~~
S=\sin \omega. \end{array} \eqno (16) $$

\noindent
According to the definition, we have [18]
$$R(\alpha, \beta, \gamma)=\left(\begin{array}{ccc}
c_{\alpha}c_{\beta}c_{\gamma}-s_{\alpha}s_{\gamma}
&~~-c_{\alpha}c_{\beta}s_{\gamma}-s_{\alpha}c_{\gamma}~~
&c_{\alpha}s_{\beta} \\
s_{\alpha}c_{\beta}c_{\gamma}+c_{\alpha}s_{\gamma}
&-s_{\alpha}c_{\beta}s_{\gamma}+c_{\alpha}c_{\gamma}
&s_{\alpha}s_{\beta} \\
-s_{\beta}c_{\gamma}&s_{\beta}s_{\gamma}&c_{\beta} \\
\end{array} \right), $$
$$\begin{array}{ll}
x_{1}+ix_{2}=r_{x}e^{i\alpha}s_{\beta},~~~
&y_{1}+iy_{2}=r_{y}e^{i\alpha}\left(c_{\beta}c_{\gamma}S+s_{\beta}C
+is_{\gamma}S\right),\\
x_{3}=r_{x}c_{\beta},~~~~~
&y_{3}=r_{y}\left(-s_{\beta}c_{\gamma}S+c_{\beta}C \right). \end{array} 
\eqno (17) $$

Through the replacement of variables:
$$\begin{array}{l}
(r_{x}, \theta_{x}, \varphi_{x}, r_{y}, \theta_{y}, \varphi_{y})
~~\longrightarrow~~(r_{x}, r_{y}, \omega, \alpha, \beta, \gamma), \\
\alpha=\varphi_{x},~~~~~~~~~~\beta=\theta_{x}, \\
C=c_{x}c_{y}+s_{x}s_{y}\cos(\varphi_{x}-\varphi_{y}), \\
\cot \gamma=\displaystyle {s_{x}c_{y}-c_{x}s_{y}\cos(\varphi_{x}-\varphi_{y})
\over s_{y}\sin (\varphi_{x}-\varphi_{y}) },
\end{array} \eqno (18) $$

\noindent
we obtain
$$\begin{array}{rl}
\bigtriangleup_{{\bf x}}&=~\displaystyle {1\over r_{x}}
\displaystyle {\partial^{2} \over \partial r_{x}^{2}}
r_{x} +\cdots ,\\[3mm]
\bigtriangleup_{{\bf y}}&=~\displaystyle {1\over r_{y}}
\displaystyle {\partial^{2} \over \partial r_{y}^{2}}
r_{y} +\displaystyle {1\over r_{y}^{2}S}
\displaystyle {\partial \over \partial \omega}
S \displaystyle {\partial \over \partial \omega} +\cdots ,
\end{array} \eqno (19) $$

\noindent
where the neglected terms are those differential terms only with 
respect to the rotational variables $\alpha$, $\beta$ and $\gamma$. 
Now, 
$$\begin{array}{l}
\left.\displaystyle {\hbar^{2} \over 2M} 
\left\{ \displaystyle {1\over r_{x}}
\displaystyle {\partial^{2} \over \partial r_{x}^{2}}
r_{x} + \displaystyle {1\over r_{y}}
\displaystyle {\partial^{2} \over \partial r_{y}^{2}}
r_{y} +\displaystyle {1\over r_{y}^{2}S}
\displaystyle {\partial \over \partial \omega}
S \displaystyle {\partial \over \partial \omega} \right\}
\Psi_{\ell \ell \lambda}({\bf x,y})\right|_{\xi_{j}=const.}\\
\left.~~~~~=E \Psi_{\ell \ell \lambda}({\bf x,y})\right|_{\xi_{j}=const.}, 
\end{array} \eqno (20) $$

\noindent
where $\Psi_{\ell \ell \lambda}({\bf x,y})$ is given in Eq. (14).

Through a direct calculation, we obtain
$$\begin{array}{l}
\displaystyle {1\over r_{x}}
\displaystyle {\partial^{2} \over \partial r_{x}^{2}}
r_{x}Q^{\ell\lambda}_{q}({\bf x,y})
=\displaystyle {q(q+1)\over r_{x}^{2}}
Q^{\ell\lambda}_{q}({\bf x,y}), \\[3mm]
\displaystyle {1\over r_{y}}
\displaystyle {\partial^{2} \over \partial r_{y}^{2}}
r_{y} Q^{\ell\lambda}_{q}({\bf x,y})
=\displaystyle {(\ell-q+\lambda)(\ell-q+\lambda+1)\over r_{y}^{2}}
Q^{\ell\lambda}_{q}({\bf x,y}), \\[4mm]
\displaystyle {1\over r_{y}^{2}S}
\displaystyle {\partial \over \partial \omega}
S \displaystyle {\partial \over \partial \omega} 
Q^{\ell\lambda}_{q}({\bf x,y})\\[3mm]
~~~=\left\{ (\ell-q) \left[(\ell-q+2\lambda)\cot^{2} \omega -1\right]
\right.\\[2mm]
\left.~~~~~~+\lambda \left(\cot^{2} \omega -1\right) \right\}
 Q^{\ell\lambda}_{q}({\bf x,y})/r_{y}^{2} \\[2mm] 
~~~~~~-(q-\lambda+1)\left(2\ell-2q+2\lambda-1\right)\left(C/S^{2}\right)
Q^{\ell\lambda}_{q+1}({\bf x,y})/\left(r_{x}r_{y}\right)\\[2mm] 
~~~~~~+(q-\lambda+1)(q-\lambda+2)S^{-2}
Q^{\ell\lambda}_{q+2}({\bf x,y})/r_{x}^{2} .
\end{array} \eqno (21) $$

Therefore, the coefficients $f_{q}^{\ell \lambda}$ satisfies a
system of linear algebraic equations with the equation
number $(\ell-\lambda+1)$:
$$\begin{array}{l}
(2ME/\hbar^{2})f_{q}^{\ell\lambda}=\left\{q(q+1)/r_{x}^{2}
+(\ell-q+\lambda)(\ell-q+\lambda+1)/r_{y}^{2} \right. \\
\left.~~~~~+\left[(\ell-q)(\ell-q+2\lambda)\cot^{2} \omega -(\ell-q)
+\lambda \left(\cot^{2} \omega -1\right) \right]/r_{y}^{2}
\right\}f_{q}^{\ell\lambda}\\[1mm]
~~~~~-\left\{(q-\lambda)\left(2\ell-2q+2\lambda+1\right)C/\left(S^{2}
r_{x}r_{y}\right)\right\}
f^{\ell\lambda}_{q-1}\\ [2mm]
~~~~~+\left\{(q-\lambda)(q-\lambda-1)/\left(S^{2}r_{x}^{2}\right)
\right\}f^{\ell\lambda}_{q-2} .
\end{array} \eqno (22) $$

\noindent
where $r_{x}$, $r_{y}$ and $\omega$ are constant.

Due to the spherical symmetry, the energy level 
with the given total angular momentum $\ell$ is $(2\ell+1)$-degeneracy 
(normal degeneracy). Furthermore, since $\lambda \leq q \leq \ell$, 
there are $(\ell+1)$ sets of solutions with the parity $(-1)^{\ell}$ 
and $\ell$ sets of solutions with the parity $(-1)^{\ell+1}$. This 
conclusion coincides with that by Wigner (see P. 218 in [17]).
When $\ell=0$ we have the constant solution with zero energy and
even parity. When $\ell=1$, we have one set of solutions 
$\Psi_{\ell m1}$ with the even parity and two sets of solutions 
$\Psi_{\ell m0}$ with the odd parity:
$$\begin{array}{l}
\Psi_{111}({\bf x,y})=(x_{1}+ix_{2})y_{3}-
x_{3}(y_{1}+iy_{2}),\\[1mm]
E_{11}=\hbar^{2}/\left(Mr_{x}^{2}\right)
+\hbar^{2}/\left(2Mr_{y}^{2}\sin^{2}\omega\right), \\[1mm]
\Psi_{110}^{(1)}=x_{1}+ix_{2},~~~~~
E_{10}^{(1)}=\hbar^{2}/\left(Mr_{x}^{2}\right), \\[1mm]
\Psi_{110}^{(2)}=\displaystyle {C \over S^{2}r_{x}r_{y}}
(x_{1}+i x_{2})+\left(\displaystyle {2 \over r_{x}^{2}}
-\displaystyle {1 \over S^{2} r_{y}^{2}} \right) (y_{1}+iy_{2}),\\[1mm]
E_{10}^{(2)}=\hbar^{2}/\left(2Mr_{y}^{2}\sin^{2}\omega\right), 
\end{array} \eqno (23) $$

\noindent
It is similar to obtain the solutions with the higher orbital
angular momentum $\ell$. The partners of the solutions with the 
smaller eigenvalues of $L_{3}$ can be calculated from them by 
the lowering operator $L_{-}$. 

\vspace{4mm}
\noindent
{\bf 4. CONCLUSION}

\vspace{2mm}
\noindent
In summary, we have reduced the three-body Schr\"{o}dinger equation
for any given total orbital angular momentum and parity
to a system (12) of the coupled partial differential equations with 
respect only to three variables, describing the internal degrees
of freedom in a three-body problem. This equation system is
a good starting point for solving the quantum three-body problems. 
As an example, we obtain the exact solution to the Schr\"{o}dinger 
equation for the rigid body.

\vspace{8mm}
\noindent
{\bf Acknowledgements}. The author would like to 
thank Prof. Hua-Tung Nieh and Prof. Wu-Yi Hsiang for drawing 
his attention to the quantum three-body problems. This work 
was supported by the National Natural Science Foundation of 
China and Grant No. LWTZ-1298 of the Chinese Academy of Sciences.

\vspace{6mm}
\noindent
{\bf REFERENCES}

\vspace{2mm}
1. \parbox[t]{4.4in}{L. D. Faddeev, {\it Sov. Phys. JETP} {\bf 12}, 
1014 (1961); {\it Sov. Phys. Dokl}. {\bf 6}, 384 (1961);
{\it Sov. Phys. Dokl}. {\bf 7}, 600 (1963).}

\vspace{2mm}
2. \parbox[t]{4.4in}{N. Barnea and V. Mandelzweig, {\it Phys. Rev}. C 
{\bf 49}, 2910 (1994). }

\vspace{2mm}
3. \parbox[t]{4.4in}{I. L. Hawk and D. L. Hardcastle, 
{\it Comp. Phys. Commun}. {\bf 16}, 159 (1979).}

\vspace{2mm}
4. \parbox[t]{4.4in}{F. S. Levin and J. Shertzer, {\it Phys. Rev}. A 
{\bf 32}, 3285 (1985).}

\vspace{2mm}
5. \parbox[t]{4.4in}{Y. K. Ho, {\it Phys. Rev}. A {\bf 34}, 4402 (1986).}

\vspace{2mm}
6. \parbox[t]{4.4in}{I. G. Fray and B. J. Howard, {\it Chem. Phys}. 
{\bf 111}, 33 (1987). }

\vspace{2mm}
7. \parbox[t]{4.4in}{J. Z. Tang, S. Watanabe, and M. Matsuzawa, 
{\it Phys. Rev}. A {\bf 46}, 2437 (1992). }

\vspace{2mm}
8. \parbox[t]{4.4in}{B. Zhou, C. D. Lin, J. Z. Tang, S. Watanabe, 
and M. Matsuzawa, {\it J. Phys}. B {\bf 26}, 2555 (1993); 
B. Zhou and C. D. Lin, {\it J. Phys}. B {\bf 26}, 2575 (1993). }

\vspace{2mm}
9. \parbox[t]{4.4in}{M. I. Haftel and V. B. Mandelzweig, 
{\it Phys. Lett}. A {\bf 120}, 232 (1987). }

\vspace{2mm}
10. \parbox[t]{4.4in}{M. I. Haftel and V. B. Mandelzweig, 
{\it Ann. Phys}. (N.Y.) {\bf 189}, 29 (1989). }

\vspace{2mm}
11. \parbox[t]{4.4in}{R. Krivec and V. B. Mandelzweig, 
{\it Phys. Rev}. A {\bf 42}, 3779 (1990).}

\vspace{2mm}
12. \parbox[t]{4.4in}{F. M. Lev, {\it Fortschritte der Physik}, 
{\bf 31}, 75 (1983).}

\vspace{2mm}
13. \parbox[t]{4.4in}{H. Letz, {\it Nuovo Cimento} B {\bf 26}, 522 (1975).}

\vspace{2mm}
14. \parbox[t]{4.4in}{E. F. Redish, "Lectures in the Quantum 
Three-Body Problem", Preprint MDDP-TR-77-060, 1977.}

\vspace{2mm}
15. \parbox[t]{4.4in}{J. Ginibre and M. Moulin, "Hilbert Space 
Approach to the Quantum Mechanical Three-Body Problem", Preprint 
LPTHE-TH 74/8, 1974.}

\vspace{2mm}
16. \parbox[t]{4.4in}{R. Krivec, {\it Few-Body Systems}, {\bf 25}, 
199 (1998) and references therein.}

\vspace{2mm}
17. \parbox[t]{4.4in}{E. P. Wigner, "Group Theory and its Application 
to the Quantum Mechanics of Atomic Spectra" (Academic Press, New 
York 1959).}

\vspace{2mm}
18. \parbox[t]{4.4in}{A. R. Edmonds, "Angular Momentum in Quantum 
Mechanics" (Princeton University Press, 1957).}

\vspace{2mm}
19. \parbox[t]{4.4in}{W. T. Hsiang and W. Y. Hsiang, "On the 
reduction of the Schr\"{o}dinger's equation of three-body problem 
to a system of linear algebraic equations", Preprint, 1998.}

\vspace{2mm}
20. \parbox[t]{4.4in}{Zhong-Qi Ma and An-Ying Dai, "Quantum 
three-body problem", Preprint, physics/9905051, 1999.}

\end{document}